# A BaTiO$_3$-based electro-optic Pockels modulator monolithically integrated on an advanced silicon photonics platform

Felix Eltes, Christian Mai, Daniele Caimi, Marcel Kroh, Youri Popoff, Georg Winzer, Despoina Petousi, Stefan Lischke, J. Elliott Ortmann, Lukas Czornomaz, Lars Zimmermann, Jean Fompeyrine, Stefan Abel

*Abstract*—To develop a new generation of high-speed photonic modulators on silicon-technology-based photonics, new materials with large Pockels coefficients have been transferred to silicon substrates. Previous approaches focus on realizing stand-alone devices on dedicated silicon substrates, incompatible with the fabrication process in silicon foundries. In this work, we demonstrate monolithic integration of electro-optic modulators based on the Pockels effect in barium titanate (BTO) thin films into the back-end-of-line of a photonic integrated circuit (PIC) platform. Molecular wafer bonding allows fully PIC-compatible integration of BTO-based devices and is, as shown, scalable to 200 mm wafers.

The PIC-integrated BTO Mach-Zehnder modulators outperform conventional Si photonic modulators in modulation efficiency, losses, and static tuning power. The devices show excellent $V_\pi L$ (0.2 Vcm) and $V_\pi L\alpha$ (1.3 VdB), work at high speed (25 Gbps), and can be tuned at low static power consumption (100 nW). Our concept demonstrates the possibility of monolithic integration of Pockels-based electro-optic modulators in advanced silicon photonic platforms.

## I. INTRODUCTION

SILICON technology based photonic integrated circuits (Si PIC) are becoming essential for various applications in the domain of communication technologies [1]. For large data centers, Si PIC technology offers attractive features for transceivers targeting intra and inter data-center communication. Utilizing advanced manufacturing techniques for the co-integration of optics and electronics enables high-speed and cost-effective transceiver solutions that take advantage of device scaling opportunities. Targeting a truly monolithic integration with CMOS (or Bi-CMOS) is crucial for such transceivers. The co-use of the back-end of line (BEOL) by photonic and electronic devices results in the smallest possible parasitics, which is a pre-requisite for efficient RF driving. Standard Si PIC modulators are based on phase shifters using the free-carrier dispersion effect. In terms of phase-shifter properties this is not the optimum solution. Besides a rather low modulation efficiency, nonlinearity and high-loss also limit modulator performance. The impossibility of disentangling amplitude and phase modulation also restricts their use for higher modulation formats [2], [3]. In addition, the high junction capacitance limits the achievable bandwidth [3] and is detrimental for power consumption. It is therefore highly desirable to enable - in a silicon photonic technology - pure electro-optic phase shifters exploiting the Pockels effect, in order to provide a solution without residual amplitude modulation, yet with high linearity, high efficiency and low optical loss. Recently, this field of research experienced a renaissance, with several attempts to demonstrate Pockels modulators potentially compatible with Si-PIC. Different strategies are being followed, using either a strain-induced Pockels effect in silicon [4], using well-known Pockels materials such as LiNbO$_3$, bonded onto silicon by direct wafer bonding [5], [6], or introducing novel materials with large Pockels coefficients [7], [8]. All these approaches have intrinsic weaknesses, coming either from a weak Pockels effect [4], the limited availability of large wafer sizes [5], [6], thermal stability issues [8], or incompatibility with standard fabrication processes [7].

Our approach utilizes single crystalline, ferroelectric BaTiO$_3$ (BTO) as a material having a large Pockels coefficient and

This work was supported in part by the Swiss National Foundation under project 200021_159565 PADOMO, by the European Commission under grant agreement no. H2020-ICT-2015-25-688579 (PHRESCO) and H2020-ICT-2017-1-780997 (plaCMOS), and by the Swiss State Secretariat for Education, Research and Innovation under contract no. 15.0285.

F. Eltes, D. Caimi, Y. Popoff, L. Czornomaz, J. Fompeyrine, and S. Abel are with IBM Research – Zurich, Säumerstrasse 4, 8803 Rüschlikon, Switzerland (e-mail: fee@zurich.ibm.com, cai@zurich.ibm.com, ypo@zurich.ibm.com, luk@zurich.ibm.com, jfo@zurich.ibm.com, sab@zurich.ibm.com).

Y. Popoff is with EMPA, 8600 Dübendorf, Switzerland.

M. Kroh was with IHP, Im Technologiepark 25, 15236 Frankfurt (Oder), Germany. He is now with Silicon Radar GmbH, Im Technologiepark 1, 15236 Frankfurt (Oder), Germany (e-mail: marcel.kroh@siliconradar.com).

C. Mai, G. Winzer, D. Petousi, S. Lischke, L. Zimmermann are with IHP, Im Technologiepark 25, 15236 Frankfurt (Oder), Germany (e-mail: cmai@ihp-microelectronics.com, winzer@ihp-microelectronics.com, petousi@ihp-microelectronics.com, lischke@ihp-microelectronics.com, lzimmermann@ihp-microelectronics.com).

L. Zimmermann is with Technische Universitaet Berlin, FG Si-Photonik, Einsteinufer 25, 10587 Berlin, Germany.

J. E. Ortmann is with the Department of Physics, The University of Texas, Austin, Texas 78712, United States (e-mail: jortmann@utexas.edu).

where an integration path of single crystalline layers with silicon does exist [9]–[11]. Over the past years, great progress has been made in developing a hybrid BTO/silicon technology, including passive structures with low-propagation losses [12], active electro-optic switching [13], [14], excellent thermal stability [15], and, very recently, large Pockels coefficients of $r_{42}$ = 923 pm/V and high-speed modulation in photonic devices[10], [11]. However, previous work was developed on silicon-on-insulator substrates – without attention to process integration in a standard PIC or electronic PIC (EPIC) process. Here, we overcome this limitation and demonstrate the integration of highly efficient BTO Pockels modulators in the BEOL of a silicon photonic process flow and show the scalability of our approach up to 200 mm, making this an attractive technology for high-speed transceivers.

## II. Technology Concept

Our concept of high-speed transceivers relies on the monolithic integration of BTO thin films via direct wafer bonding above an interlayer dielectric (ILD) in a standard EPIC flow (Fig. 1a) [16]. The bonding step can be performed on top of any ILD above the front-end-of-line (FEOL) structures. Using wafer bonding we can first deposit BTO epitaxially on a silicon substrate, and then transfer the epitaxial layer onto an amorphous substrate, such as an ILD. Having an epitaxial BTO film is of importance for two reasons: First, the low defectivity in single-crystalline films is crucial for achieving a large effective Pockels effect in the material [17]. Second, the low surface roughness of epitaxial BTO films is critical for obtaining high bonding yield. To fabricate BTO thin films, we use a deposition process based on molecular beam epitaxy [9], [18], which relies on the epitaxial growth on Si wafers and can thus be scaled to large wafer sizes. The availability of large substrates is major a benefit compared to the bonding of $LiNbO_3$ on silicon or to the epitaxial growth on crystalline oxides, both approaches being limited by the available substrate or donor crystal sizes.

The BTO devices are based on a strip-loaded waveguide geometry, where a Si strip on top of BTO guides the optical mode (Fig. 1b). Lateral electrodes for phase shifters are made using metal lines fabricated in the top metal level of the BEOL before BTO integration, combined with a final metallization after BTO integration. Optical simulations are used to inform the design of the BTO-Si waveguides and to ensure substantial overlap of the transverse electric (TE) optical mode and the BTO layer at a wavelength of 1550 nm. Using a 170-nm-thick BTO layer loaded with a 100-nm-thick Si strip, we achieve an optical overlap of 38% between the first order TE mode and the BTO layer. The BTO/Si phase shifters can be used in Mach-Zehnder modulators (MZMs), or ring modulators. In this work we used unbalanced MZMs, with multi-mode interference splitters. Grating couplers were used to couple light in and out of the devices (Fig. 1c). The magnitude of the refractive index change induced by the Pockels effect is strongly anisotropic and depends on the relative orientation of the crystalline axes, the optical electrical field, and the modulating electric field [10], [19]. To ensure the maximum response we designed phase shifters with waveguides along the BTO[110] direction.

## III. Integration and Fabrication

We deposited BTO thin films on $SrTiO_3$-buffered silicon-on-insulator substrates with 100 nm top Si using a previously reported process [9], [18]. Deposition of BTO using molecular beam epitaxy ensures a high-quality single-crystal film. After BTO deposition, we transferred the BTO layer and the top Si layer onto a planarized acceptor wafer using thin alumina layers for adhesion at the bonding interface. The donor wafer was subsequently removed by a combination of mechanical grinding and chemical etching, resulting in a high transfer yield from the source wafer.

To demonstrate the scalability of our approach, we transferred BTO layers grown on a 200 mm SOI substrate onto another 200 mm silicon wafer that had been capped with a thermal oxide (Fig. 2a). The transferred $BaTiO_3$ layer was thoroughly characterized using X-ray diffraction (XRD) and ellipsometry (Fig. 2b). The out-of-plane lattice parameter and rocking curve show good crystalline homogeneity with only minor variations along the 100 mm radius. Additionally, the thickness of the BTO, measured by ellipsometry, varies only minimally across the wafer. The observed variations in thickness contain a significant uncertainty due to variations also in the other layers of the stack.

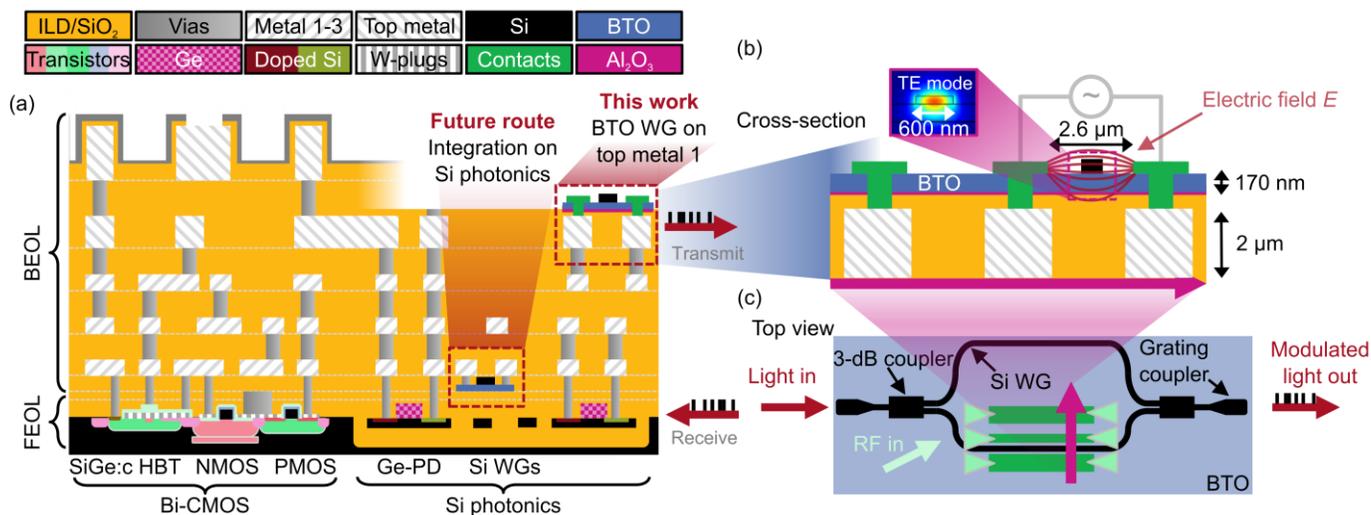

Fig. 1. Scheme for monolithic integration of BTO/Si on PIC platforms. (a) Schematic cross-sections of PIC with electrical and optical front-end, and BTO integration in the back-end (this work) or front-end (future route). (b) Cross-section of active BTO/Si waveguide used for electro-optic modulators. The electrodes (shown in grey) are fabricated in the BEOL of the PIC platform. (c) Schematic layout of BTO/Si electro-optic modulator reported in this work. The BTO/Si active waveguide is used as phase shifter in a Mach-Zehnder modulator.

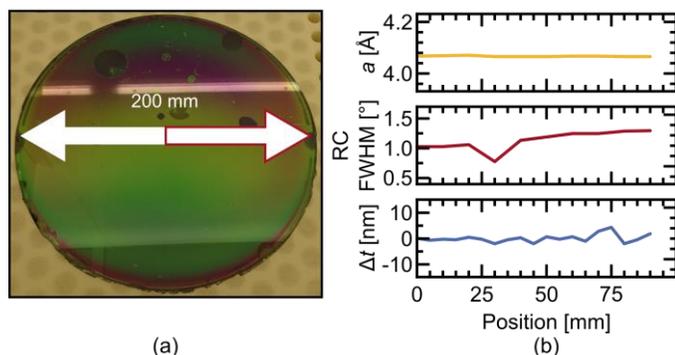

Fig. 2. Wafer bonding transfer of thin-film BTO between 200 mm source and target wafers. (a) Photo of transferred BTO layer. (b) Radial measurements of the homogeneity of the transferred BTO layer. XRD and ellipsometry was used to measure the lattice parameter, rocking curve, and thickness of the BTO, all of which show good homogeneity across the 200 mm wafer.

For the fabrication of modulators (Fig. 3), we used 200 mm target wafers, processed following a PIC flow having the same BEOL processes as EPIC runs [20]. In this work, the BEOL process of the PIC run was interrupted at the 4th metallization level, top metal 1 (TM1), after ILD planarization. We transferred a 170-nm-thick BTO layer from a 50 mm SOI wafer onto the planarized PIC wafer. Si waveguides were patterned by dry etching. In order to ensure a homogenous electric field across the BTO and to avoid a voltage drop over the thin ILD layer between BTO and TM1, vias to TM1 were etched through the BTO and the ILD along the waveguides. With a final metallization step, we extended the buried RF lines on top of the BTO. A cross-sectional electron micrograph (Fig. 4) demonstrates the successful fabrication of BTO/Si modulators on the PIC substrate.

Direct wafer bonding using $Al_2O_3$ adhesion layers has a temperature budget well within the limits of the BEOL process [21]. However, annealing steps at temperatures up to 350°C are needed to reduce the propagation losses in the BTO layer [12].

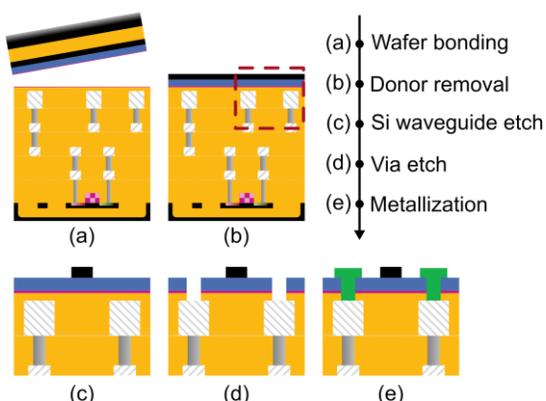

Fig. 3. Simplified process flow for integration of BTO modulators in the BEOL of a PIC process flow. Schematics of the cross-sections (left) are shown at various steps in the process (right). After wafer-bonding of the BTO and Si layers, the Si is patterned into a strip-loaded waveguide, after which vias and contacts are fabricated. Figures (c)-(d) show a magnification of the region within the dashed rectangle in (b).

It is therefore necessary to verify that the BTO integration does not cause any degradation of FEOL components. As the Ge photodiodes fabricated in the FEOL are highly sensitive to thermal degradation, we characterized their performance before and after BTO device integration. We cannot detect any degradation in either dark current or high-speed signal detection performance (Fig. 5). The absence of such degradation confirms that our integration strategy is compatible with the thermal limitations of the FEOL and BEOL processes, making integration of BTO devices compatible with PIC platforms, and fulfills the prerequisites for compatibility with EPIC platforms.

IV. DEVICE PERFORMANCE

To characterize the device performance, we used both passive ring resonators and active MZMs. The ring resonators had a radius of 30 μm, to ensure negligible bending losses, and allow accurate extraction of propagation loss. From the high Q-factor (~50,000) of ring resonators we extract a propagation loss of 5.8 dB/cm. Since the BTO layer itself has only minimal contributions to the propagation losses [12], we are instead limited by scattering losses in the Si waveguides. Using an optimized patterning process the propagation losses can be reduced further. We performed electro-optic characterization on unbalanced MZMs with phase shifter lengths of 1-2 mm. By applying a voltage to one arm of the MZM and recording the induced phase shift as a function of the applied voltage (Fig. 6), we extracted the DC $V_\pi L$ value as 0.23 Vcm. This value is 10 times smaller than state-of-the-art Si depletion-type plasma-dispersion modulators ($V_\pi L$~2 Vcm) [3, 17] and comparable to integrated silicon semiconductor-insulator-semiconductor capacitor (SISCAP) modulators ($V_\pi L$ of ~0.2 Vcm) [23]. When taking into account propagation losses $\alpha$ we reach a $V_\pi L\alpha$ of 1.3 VdB, which is significantly better than any available high-speed Si modulator ($V_\pi L\alpha$ >10 VdB). In the current devices the propagation losses are limited by scattering from roughness in the Si waveguide. By improved processing, propagation losses can be reduced to less than 3 dB/cm, resulting in a $V_\pi L\alpha$ < 0.7 VdB. The low $V_\pi L\alpha$ shows one of the key advantages of the hybrid BTO/Si technology compared to alternative modulator concepts: BTO/Si shows both a large electro-optic response and low insertion losses, since neither high doping levels nor absorbing materials are needed in the modulator design.

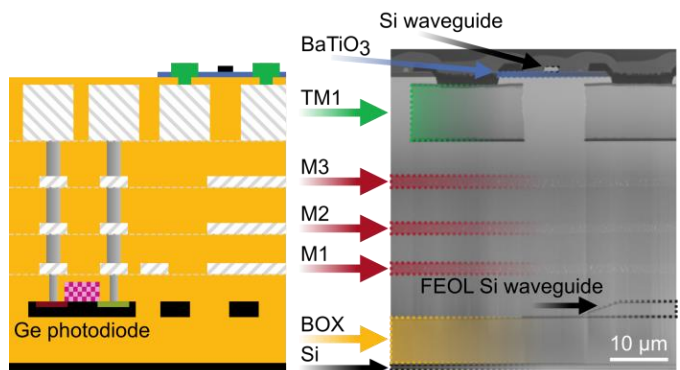

Fig. 4. Cross-sectional STEM image of BTO modulator integrated after top metal 1 (TM1) in BEOL process of a Si PIC wafer. The schematic shows how the modulator was integrated in this work. The electron micrograph shows the successful integration of BTO/Si modulators. Intermediate metal levels (M1 to M3) as well as the FEOL levels can be identified.

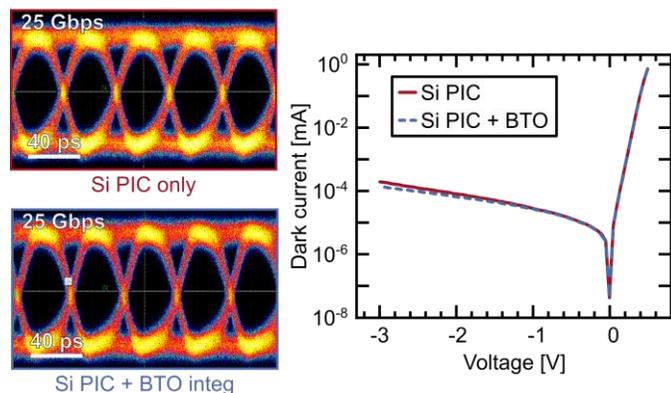

Fig. 5. Comparison of FEOL Ge photodiode performance before and after integration BTO modulators. The photodiodes were characterized by recording a modulated data signal, and by measuring the dark-current. No detectable degradation is caused by integration of BTO modulators, showing that the integration scheme is compatible with the PIC FEOL.

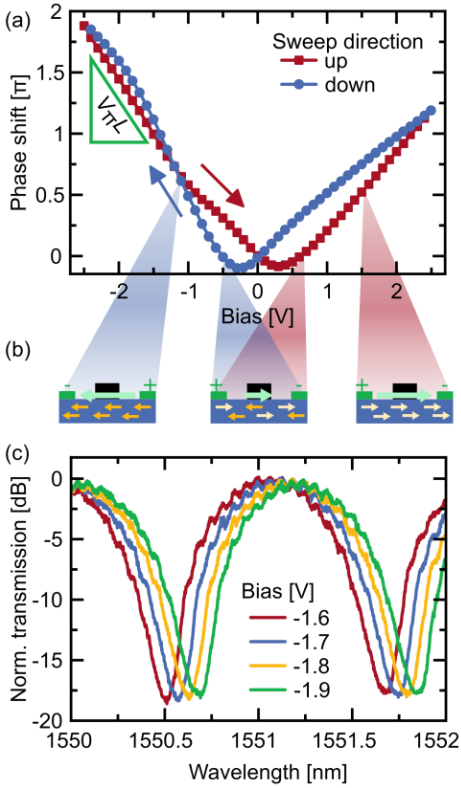

Fig. 6. (a) Induced phase shift when applying voltage to one arm of a 2-mm-long MZM. The response is linear at large voltages but shows non-linear, hysteretic contributions for small biases due to ferroelectric domain switching in the BTO layer as visualized in (b): The yellow arrows correspond to the polarization of ferroelectric domains. In the middle configuration the EO response of oppositely oriented domains cancel out, resulting in a vanishing effective Pockels coefficient. A sufficient bias voltage can align the domains to maximize the EO response. The ferro-electric domain switching is a slow effect that does not occur at frequencies >1 GHz. (c) Transmission spectra at various bias voltages in the poled regime used for $V_\pi L$ extraction.

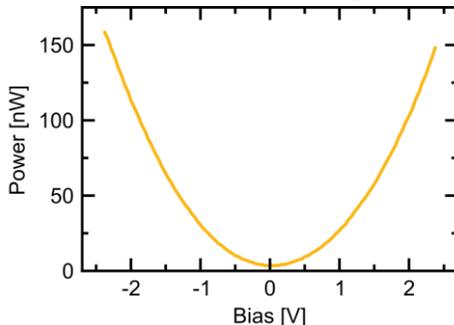

Fig. 7. Power-voltage characteristics of a 2-mm-long MZM device. The leakage current is small in the full bias range (<100 nA), resulting in low static power consumption of ~100 nW at the operating point of 2 V.

When sweeping the bias voltage, the phase shift of the MZM (Fig. 6a) exhibited a hysteretic behavior, consistent with the ferroelectric nature of BTO [10]. The hysteresis curve illustrates the need for poling the BTO layer with a bias above the coercive field (~ 1 V) to maximize the electro-optic response. For smaller bias voltages, mixed ferroelectric domain states result in a reduced effective Pockels effect $r_{eff}$, which ultimately vanishes for evenly populated domain states [9] (Fig. 6b). The cancellation effect of opposing domains causes a deviation from the linear phase response when varying a DC voltage: The total electro-optic response is the convolution of the linear Pockels effect and nonlinear domain switching effects. To isolate the Pockels effect from the electro-optical response we extract the $V_\pi$ at the extremes of the curve shown in Fig. 6a, where all domains have been poled. The re-orientation of ferroelectric domains is a relatively slow process (<<1 GHz) [24], which does not impact the operation of the modulator at high frequency – even at a bias below the poling voltage.

Moreover, as the Pockels effect is an electric-field effect, very low-power tuning of the MZMs is possible. The low leakage results in extremely low tuning powers, $P_\pi$ <100 nW (Fig. 7), compared to silicon thermo-optic tuning elements which typically have a $P_\pi$ >1 mW [25]. As the Pockels-effect is a linear EO effect the device bias can be used for tuning without changing propagation losses and without affecting the modulation efficiency.

From the measured $V_\pi L$ it is possible to extract the effective Pockels coefficient $r_{eff}$ of the BTO layer using eq. (1)

$$r_{\text{eff}} = \frac{\lambda g}{n_{\text{BTO}}^3 \Gamma_{\text{BTO}} V_\pi L} \quad (1)$$

as $r_{\text{eff}}$ = 380 pm/V. Here, $\lambda$ is the operating wavelength of 1.55 µm, $g$ is the electrode-gap (2.6 µm), $n_{\text{BTO}}$ is the refractive index of BTO (2.29) as measured by ellipsometry on similar films, and $\Gamma_{\text{BTO}}$ is the EO interaction factor which can be estimated as the optical overlap with BTO (38%) assuming a homogenous electric field across the BTO. The magnitude of the extracted $r_{\text{eff}}$ is in the range of expected values for BTO thin films: The electro-optic response exceeds values previously reported on MBE-grown BTO layers on silicon [9], [13], [17], but is smaller than those reported in ref. [10], where BTO films of very high crystalline quality with rocking curves of 0.3° are reported. The variation of the magnitude of the Pockels coefficients in similar material stacks is in agreement with the dependence of the electro-optic response on the actual crystalline quality and film morphology [17].

To determine the high-frequency response, small-signal electro-optic $S_{21}$ measurements were performed on a MZM with 1-mm-long electrodes (Fig. 8). The 3-dB bandwidth is 2 GHz.

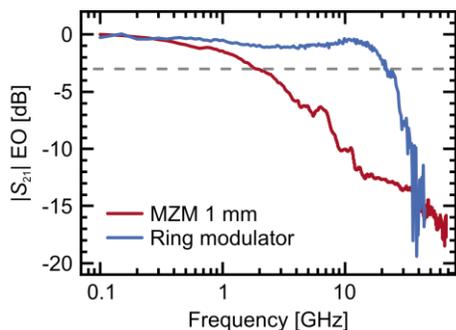

Fig. 8. Small signal frequency response of a 1-mm-long MZM, and a ring modulator with radius 10 µm. The bandwidth of the MZM is limited by mismatch between the optical and RF modes, whereas the ring modulator is limited by the photon lifetime.

The reason for this limited bandwidth is a mismatch between the optical mode and the electrical mode in the travelling wave electrodes, caused by the high dielectric constant of the BTO. Our device designs are based on moderate BTO permittivity values of $\varepsilon_{BTO}$ = 100 [26], which turned out to be strongly underestimated compared to recent reports of $\varepsilon_{BTO}$ as high as 3000 in epitaxial BaTiO$_3$ thin films [10], [27], [28]. To improve the bandwidth, the electrodes should be designed based on the actual properties of the BTO layer to achieve mode matching between the RF and optical modes. To show that the bandwidth is not limited by the electro-optic properties of the material but rather by the electrical design, we measured the bandwidth of a ring modulator with a 10 µm radius, the small radius induces bending losses resulting in a reduced Q-factor of ~15,000. The measured bandwidth of ~20 GHz (Fig. 8) is limited by the photon lifetime (Q-factor ~15,000) but demonstrates the potential for high bandwidth operation using BTO/Si devices – as confirmed in previous reports [10], [15].

We further characterized the high-speed performance of the BTO/Si modulators with data-modulation experiments using a 1-mm-long MZM. An electrical pseudorandom binary sequence (PRBS) was generated with a bit-pattern generator, without pre-emphasis or any other signal processing. The signal was amplified ($V_p$ ~2 V) and was then applied to one arm of the MZM along with a 2 V DC bias. The MZM was operated in a travelling wave configuration with an off-chip 50 Ω termination. The modulated optical signal was amplified (to compensate losses from grating couplers and from the experimental setup) and directly detected using a high-speed photodiode. Eye-diagrams were recorded on a sampling oscilloscope with 10, 20, and 25 Gbps data rates (Fig. 9). Non-closed eyes can be achieved even at 25 Gbps, however the result of the limited EO bandwidth of the modulator is qualitatively visible as a reduction eye opening from 10 to 25 Gbps. With an adapted electrode design, we expect to reach data rates >50 Gbps using MZMs.

## V. CONCLUSION

We have shown how a material (BaTiO$_3$) with the Pockels effect can be integrated into a silicon photonics platform in a scalable way using direct wafer bonding. The demonstrated Mach-Zehnder modulators show excellent performance, exceeding state-of-the-art silicon-based devices on several figures of merit, such as $V_\pi L$ and $V_\pi L\alpha$. The established integration concept provides a path for a novel generation of high-speed modulators and ultra-fast switches. The technology is however not limited to such existing components, but further enables entirely new types of devices on a silicon photonics platform. Using BTO, ultra-low-power tuning elements [14] and compact plasmonic devices [10], [29], as well as non-volatile elements for optical neuromorphic computing exploiting ferro-electric domain switching [30] are possible.

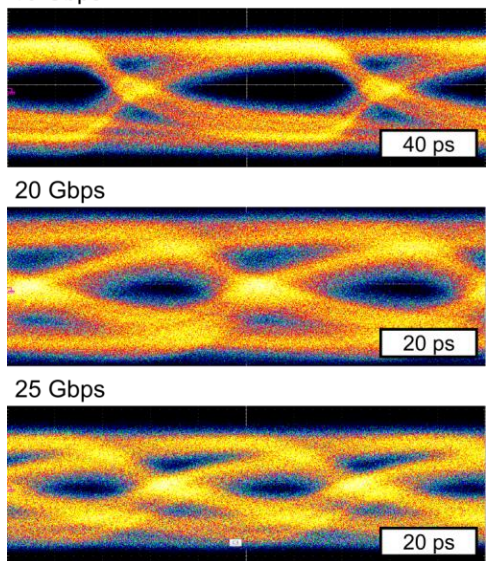

Fig. 9. Eye-diagrams from back-to-back data transmission through 1-mm-long BTO MZM in single-drive mode at 10, 20, and 25 Gbps, respectively. A bias voltage of 2 V was applied during the experiments.